\documentclass{article}
\usepackage{spconf,amsmath,epsfig}

\usepackage{graphicx}
\usepackage{times}
\usepackage[caption=false]{subfig}
\usepackage{amssymb}
\usepackage[ruled,vlined,linesnumbered]{algorithm2e}
\usepackage{epsfig}
\usepackage{multirow}
\usepackage[american]{babel}
\usepackage[pagebackref=true,breaklinks=true,colorlinks,bookmarks=false]{hyperref}
\usepackage{array}
\usepackage{enumitem}
\usepackage[table]{xcolor}
\usepackage{etoolbox}
\usepackage{threeparttable}
\usepackage{setspace}

\definecolor{Gray}{gray}{0.9}
\newcolumntype{C}[1]{>{\centering\let\newline\\\arraybackslash\hspace{0pt}}m{#1}}
\begin{document}\sloppy

\def\x{{\mathbf x}}
\def\L{{\cal L}}


\title{A Multimodal Lossless Coding Method for Skeletons in Videos}
%
\begin{spacing}{0.2}
\name{\normalsize Xiaoyi He\textsuperscript{1*}, Mingzhou Liu\textsuperscript{1*}, Weiyao Lin\textsuperscript{1\dag}, Xintong Han\textsuperscript{2}, Yanmin Zhu\textsuperscript{3}, Hongtao Lu\textsuperscript{3}, Hongkai Xiong\textsuperscript{1}\thanks{\textsuperscript{*}Equal contribution} \thanks{\textsuperscript{\dag}Corresponding Author: wylin@sjtu.edu.cn}}
\address{\normalsize \textsuperscript{1} Department of Electronic Engineering, Shanghai Jiao Tong University, China \textsuperscript{2} Malong Technologies \\
\normalsize \textsuperscript{3} Department of Computer Science and Engineering, Shanghai Jiao Tong University, China}

\maketitle
\end{spacing}

\begin{abstract}
Nowadays, skeleton information in videos plays an important role in human-centric video analysis but effective coding such massive skeleton information has never been addressed in previous work. In this paper, we make the first attempt to solve this problem by proposing a multimodal skeleton coding tool containing three different coding schemes, namely, spatial differential-coding scheme, motion-vector-based differential-coding scheme and inter prediction scheme, thus utilizing both spatial and temporal redundancy to losslessly compress skeleton data. More importantly, these schemes are switched properly for different types of skeletons in video frames, hence achieving further improvement of compression rate. Experimental results show that our approach leads to 74.4\% and 54.7\% size reduction on our surveillance sequences and overall test sequences respectively, which demonstrates the effectiveness of our skeleton coding tool.
\end{abstract}
\section{Introduction and Related work}
\label{sec:intro}
Skeleton information in videos is of increasing important recently in many applications such as event detection, video recognition, etc. For example, previous works have shown how action recognition can benefit from skeleton-based video modeling~\cite{wang2017modeling,ke2017new,tang2018deep,demisse2018pose}. A person's pose is described by multiple skeleton key joints and the skeleton information in videos represents the dynamic characteristics of body postures, which makes skeleton information widely used in human action recognition and other video analysis tasks.

Since video analysis is directly performed based on extracted features, shifting the feature extraction into the camera-integrated module can reduce the analysis server load and is highly desirable. Therefore, some feature coding methods that aim to compress and transmit different kinds of extracted features of videos are proposed recently. Duan \textit{et al.}~\cite{duan2018compact} describe the compact descriptors for video analysis, where handcrafted and deep features are compressed and transmitted in a standardized bitstream. Chen \textit{et al.}~\cite{chen2014efficient} introduce their proposed Region-of-Interest (ROI) location coding tool where the ROI location information itself is coded in the video bitstream.

Recently, reliable human skeletons can be obtained from the depth sensor using real-time skeleton estimation algorithms. However, transmitting these skeletons directly back to the analysis server is too expensive. In this paper, we argue that skeleton information in videos plays an important role in video analysis. However, existing approaches have been overlooked coding this massive skeleton information. Therefore, it is necessary to develop new algorithms to encode this skeleton data efficiently. To the best of our knowledge, this paper is the first to study coding skeleton information into bitstream.

\begin{figure}[t]
\centering
\subfloat[]{\includegraphics[height=3.6cm]{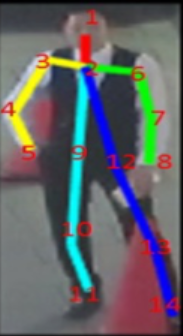}%
\label{a_ske}}
\hfil
\subfloat[]{\includegraphics[height=3.6cm]{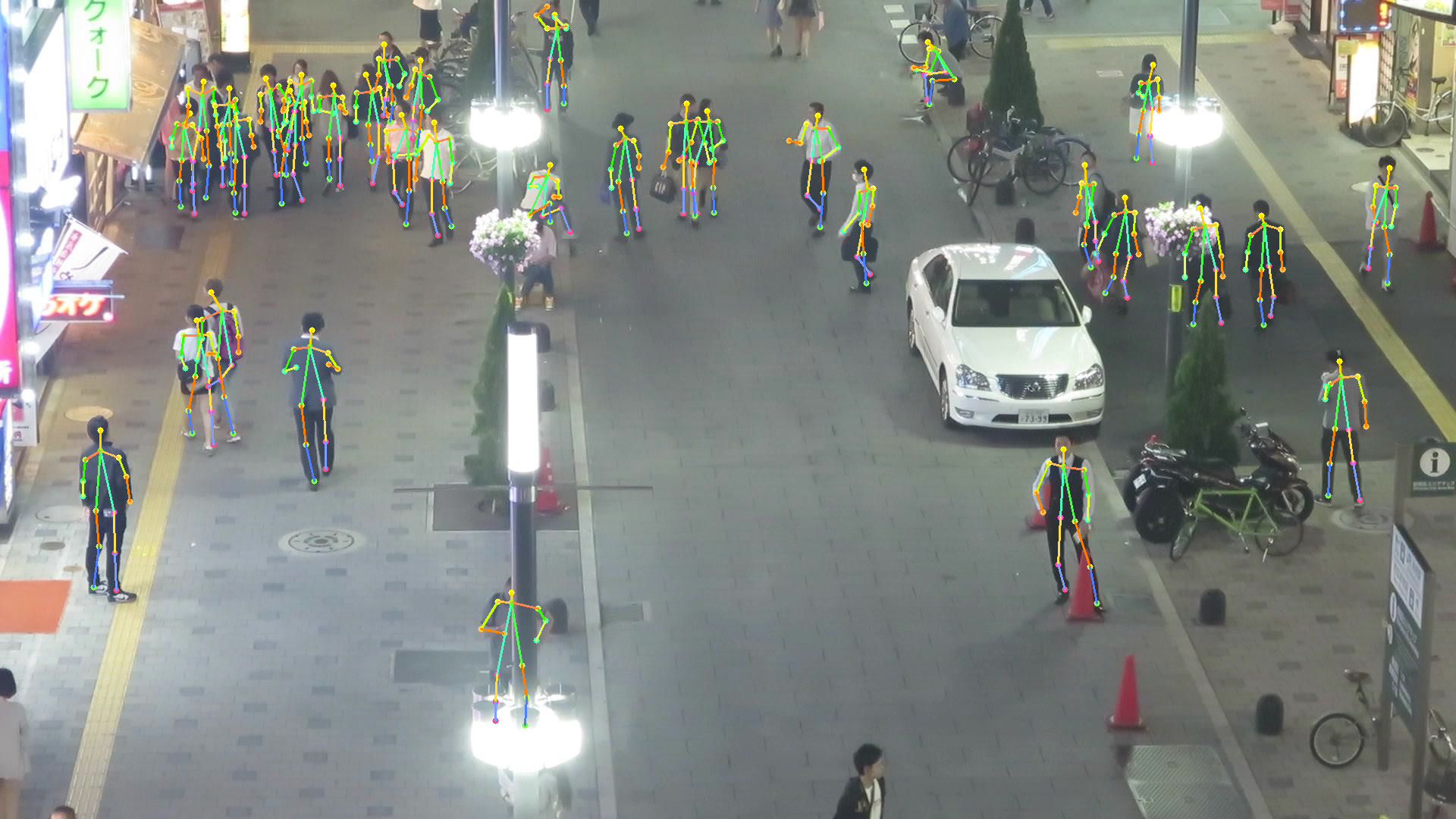}%
\label{ske_in_frame}}
\hfil
\subfloat[]{\includegraphics[width=0.8\linewidth]{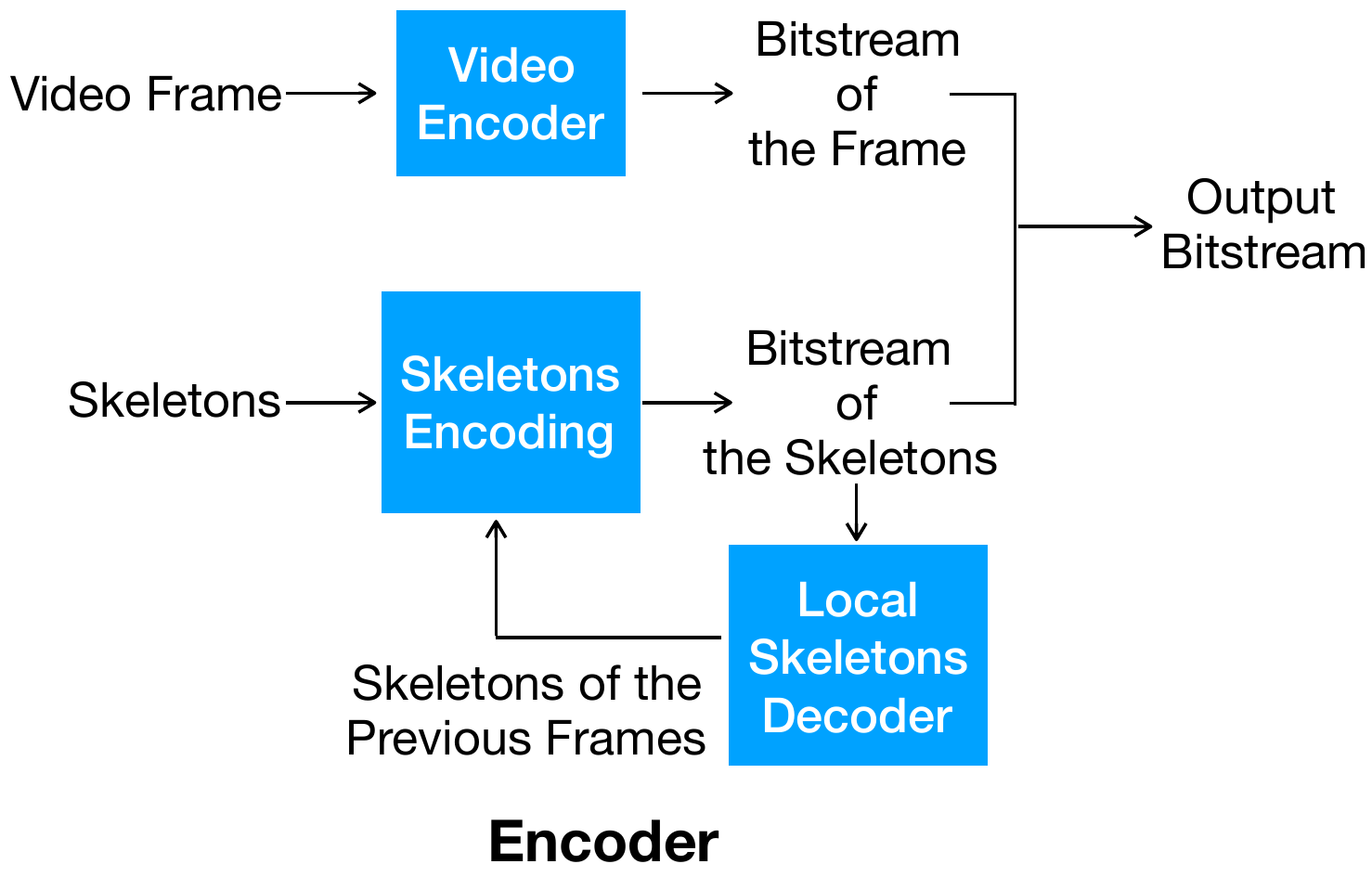}%
\label{Overview of our method}}
\caption{(a) An example of skeletons (b) Skeleton information in one video frame (c) Overview of skeletons compression algorithm}
\label{fig_sim}
\end{figure}

\begin{figure*}[t]
\centering
\includegraphics[width=0.8\linewidth]{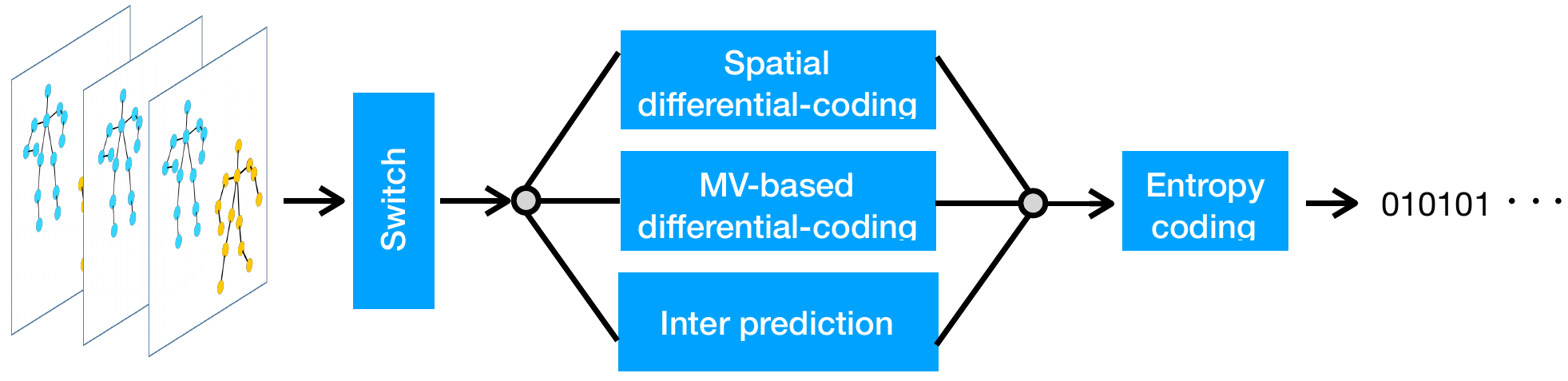}
\caption{The framework of our multimodal skeleton coding method.}
\label{fig:multi_framework}
\end{figure*}

In our case, skeletons in many video frames need to  be compressed and transmitted. We present human skeleton by fourteen key joints as shown in Fig.~\ref{a_ske}. For example, the $1^{st}$ is located at the nose and that labeled as $11^{th}$ presents the right ankle. Our task is to encode and transmit the size and location of each key point of these skeletons to the decoder. One straightforward way to do this is to directly transmit the $(x, y)$ coordinates of every key joint. This simple method can work well when there are only few people in the video. However, when the number of skeletons becomes large (for example, the video-frame shown in Fig.~\ref{ske_in_frame}), these skeleton location data will become huge and non-negligible. According to our experiments, the skeleton data will take about 42\% of the total bits for a video like Fig.~\ref{ske_in_frame} with about 35 skeletons in each frame. Therefore, new algorithms are required to efficiently compress these massive skeleton data.

To this end, we propose a novel approach to compress the skeleton information by combining skeletons encoderm lossless along with video codec, whose framework is shown in Fig.~\ref{Overview of our method}. In the encoder, the input video frame will be encoded by video encoder such as H.265. Meanwhile, the skeletons of this video frame are encoded by our skeletons encoding module that also takes the skeletons of previous frames from the local skeletons decoder as input. These previous skeletons will be used as the reference to reduce the redundancy of skeletons in the current frame. Then the resulting skeletons bitstream will be added together with the bitstream of the frame as the final output bitstream. Since the decoding process can be easily derived from the encoding process, we will only focus on discussing skeleton encoding in this paper.


The proposed multimodal skeleton coding tool contains three coding schemes: (1) Spatial differential-coding scheme, (2) Motion-vector-based (MV-based) differential-coding scheme, and (3) Inter prediction scheme, which are switched dynamically to encode different types skeletons.
In summary, our contributions are two folds:
\begin{enumerate}[nolistsep,noitemsep]
\item This is the first work to study coding skeleton information itself into bitstream. A skeleton coding tool is developed in this paper, which achieves skeletons compression in videos with up to 54.7\% compression rate on average.

\item We introduce three different schemes for skeleton coding. Furthermore, a multimodal scheme that integrates these schemes is proposed and achieves more robust skeletons encoding results.
\end{enumerate}

The rest of paper is organized as follows: Section~\ref{sec:overview} describes the framework of our skeleton information coding tool. Section~\ref{sec:detail} describes the detail of our coding tool and its three sub-schemes. Section~\ref{sec:ex} shows the experimental settings and results. Section~\ref{sec:conc} concludes this paper.

\section{Overview of our method}
\label{sec:overview}

Fig~\ref{fig:multi_framework} shows the framework of our multimodal skeleton coding algorithm. Skeletons are relayed to three coding schemes properly to achieve higher compression rate losslessly. The spatial differential-coding scheme utilizes the spatial redundancy to compress skeleton data while MV-based differential-coding scheme and inter prediction scheme are mainly based on the temporal redundancy. Thus, our multimodal skeleton coding tool can compress complex skeleton trajectories within crowed scene efficiently.

\section{The skeleton information coding tool}
\label{sec:detail}
In this section, we will first detail the definition of skeletons in video and then describe the three proposed skeleton coding schemes. Finally, a multimodal skeleton coding method is introduced.

\subsection{Definitions}
\label{def}
As we mentioned, the skeleton of a human can be described and coded by fourteen key points. According to this, we define the skeleton information as:
\begin{equation}
 \mathbf{SK_{i}} = \{l_{i}, (x_{i,1}, y_{i, 1}), (x_{i, 2}, y_{i, 2}), \dots, (x_{i, 14}, y_{i, 14})\}
\end{equation}
where $l_{i}$ is the ID of the $i^{th}$ human $\mathbf{SK_{i}}$ in one frame and $(x_{j}, y_{j})$ are the horizontal and vertical coordinates of $j^{th}$ key point of $\mathbf{SK_{i}}$ ($j \in \{0, 1, \dots, 14\}$). Note that each person has a unique ID over whole video and is decided according to its first appearing time in the video. The index of $i^{th}$ skeleton $i$ in one frame is decided according to its label. With these 29 elements, one human skeleton in one video frame can be determined uniquely.

The difference between two skeletons are defined as the set of difference between the same key joint:
\begin{equation}
 \mathbf{SK_{i} - SK_{k}} = \{(x_{i,j}-x_{k,j}, y_{i,j}-y_{k,j})|j=1,2, \dots, 14\}
\end{equation}

\begin{figure}[t]
\centering
\includegraphics[width=0.5\linewidth]{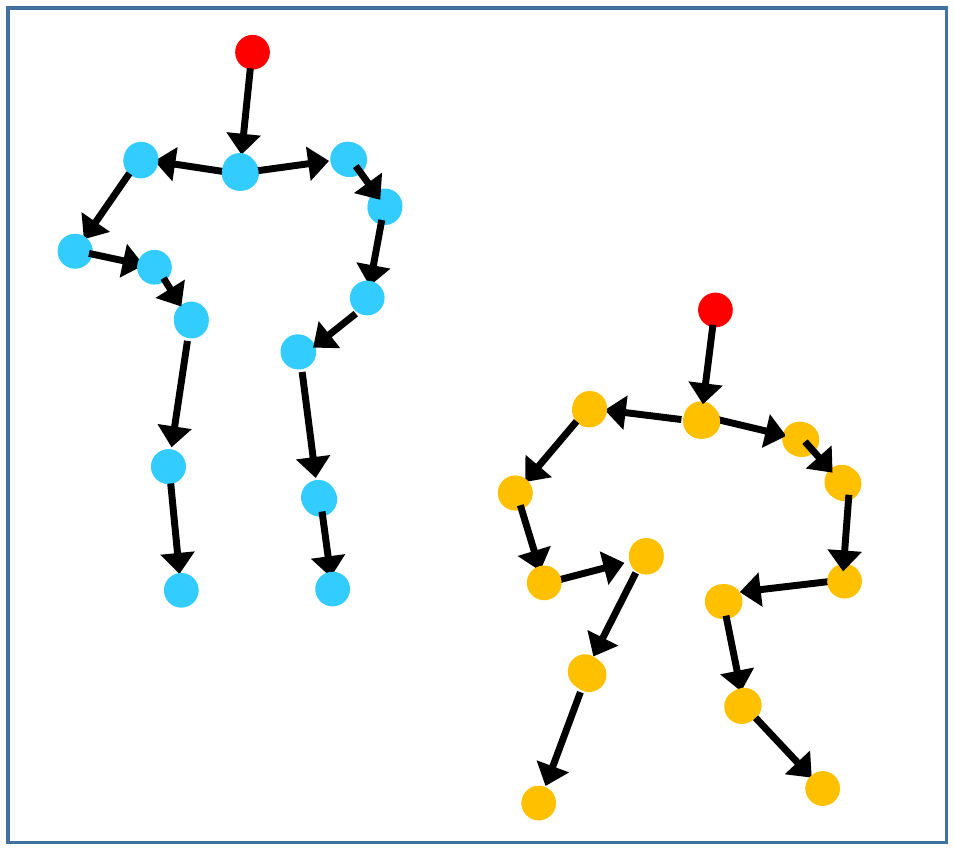}
\caption{Illustration of spatial differential-coding scheme}
\label{fig:spatial_fwk}
\end{figure}

\subsection{Skeleton Coding Schemes}
Three coding schemes are introduced in our skeleton information coding tool:

\textbf{Spatial differential-coding scheme.}\quad Considering the spatial correlation of joints within a skeleton, we developed a spatial differential-coding scheme that utilizes the spatial redundancy to compress the skeleton data.  As shown in Fig.~\ref{fig:spatial_fwk}, only the absolute coordinates of $1^{th}$ joint with the difference vectors (see the red joint and vectors between joints) of a skeleton are encoded.

The procedure is as follows: for each skeleton in a frame, the coordinates of $1^{th}$ joint are first encoded and a set $\mathbf{E}=\{1\}$ that represents the $1^{th}$ joint has been encoded is initialized. Then for each encoded joint in set $\mathbf{E}$, the difference between it and each of its neighbors are encoded. This process is repeated until all joints of a skeleton are encoded.

\begin{figure}[t]
\centering
\includegraphics[width=0.9\linewidth]{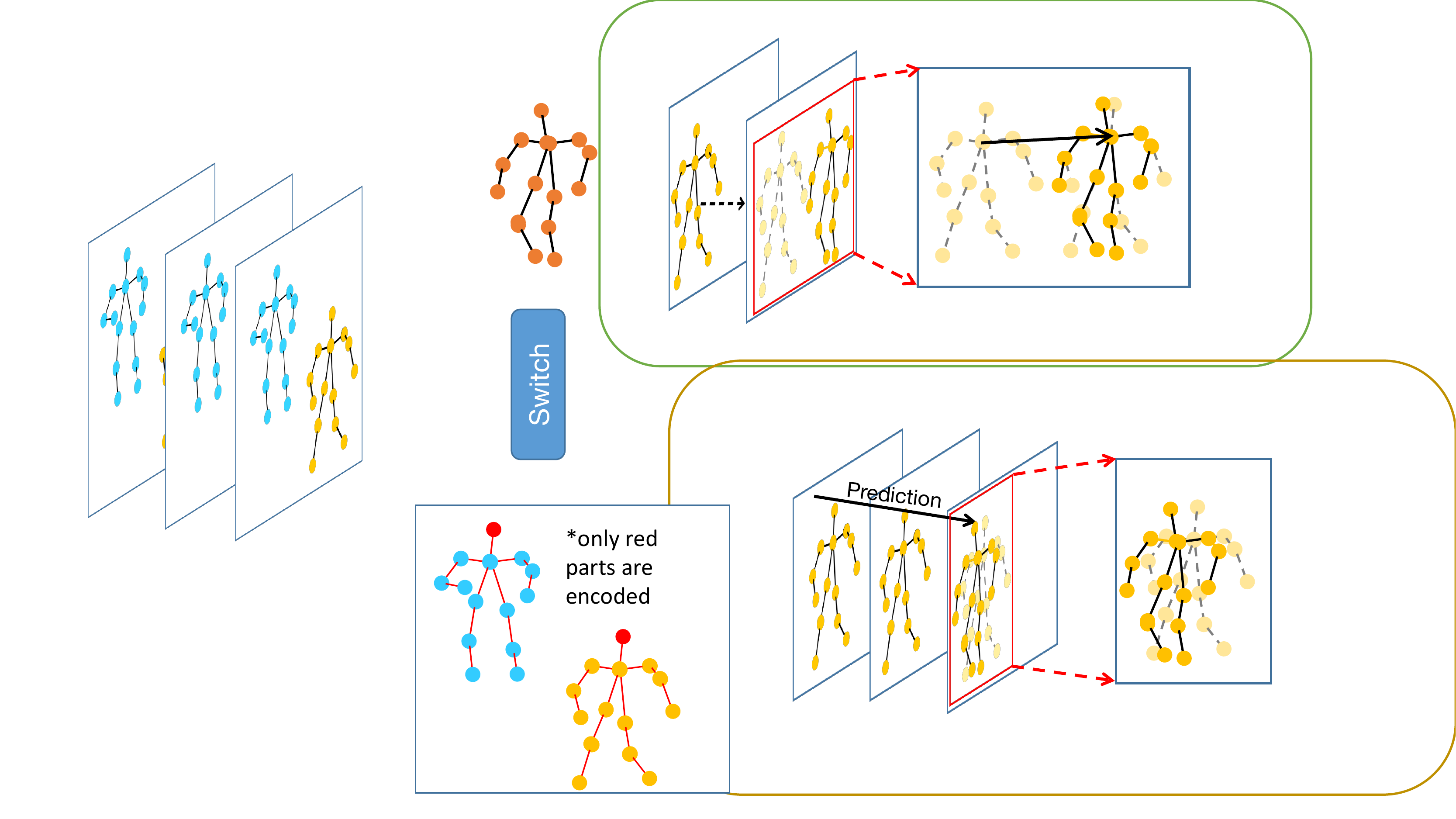}
\caption{Illustration of MV-based differential-coding scheme}
\label{fig:mv_fwk}
\end{figure}

\textbf{MV-based differential-coding scheme.}
When a lot of skeletons exist and need to be encoded in a dense crowd scene, we need a new compression algorithm for skeletons to deal with such huge amount of skeleton data efficiently. Therefore, we developed a MV-based difference-coding scheme that mainly utilizes the temporal redundancy of skeletons (the same persons' skeletons in different frames are highly correlated). As shown in Fig.~\ref{fig:mv_fwk}, the skeleton with lighter yellow joints and dash lines in $t^{th}$ frame is co-located with the one in $(t-1)^{th}$ frame. Then a predicted skeleton is obtained using the motion vector calculated with the $2^{nd}$ joint (The $2^{nd}$ joint corresponds to the center of a human) of co-located and original skeletons. Finally, the differences between the predicted skeleton using MV and the original one are encoded.

Formally, for a frame at $T=t$, the $(t-1)^{th}$ frame is chosen as the reference frame. Then for each skeleton $\mathbf{SK_{i}^{t}}$, difference between it and its corresponding skeleton in selected reference frame is encoded. More specifically, the motion vector (MV) of $2^{nd}$ joint is first calculated:
\begin{equation}
\begin{split}
  MV(\mathbf{SK_{i}}, \mathbf{SK_{k}}) &= (MV_{x}, MV_{y} ) \\
  &= (x_{i,2} - x_{k,2}, y_{i,2} - y_{k,2})
\end{split}
\end{equation}
Then the motion compensation (MC) of other joints of $\mathbf{SK_{i}}$ is achieved using the MV of $2^{nd}$ joint:
\begin{equation}
\begin{split}
  & MC(\mathbf{SK_{i}}) \\
  &  = \{(x_{i,j}+MV_{x}, y_{i,j}+MV_{y})|j = 1, 3, 4, \dots, 14\}
\end{split}
\end{equation}
Finally, the encoded parameters is defined as:
\begin{equation}
  EP(\mathbf{SK_{i}}) = \mathbf{SK_{i}} - \mathbf{ME}(\mathbf{SK_{i}})
\end{equation}

\begin{figure}[t]
\centering
\includegraphics[width=0.9\linewidth]{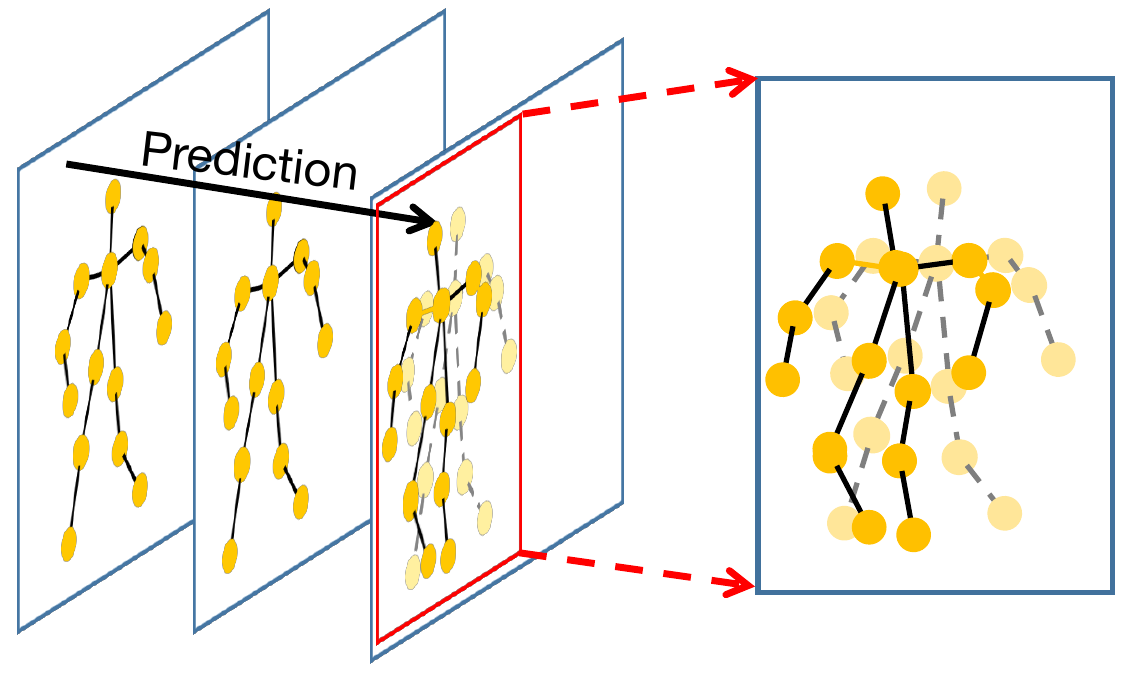}
\caption{Illustration of inter prediction scheme}
\label{fig:pre_fwk}
\end{figure}

\textbf{Inter prediction scheme.}\quad In the MV-based differential-coding scheme, the motion vector of $2^{nd}$ joint is utilized to predict all joints. It is the optimal solution when the skeleton is nearly translated from the previous to the current frame (i.e. every joint of the body moves in the same direction and over the same distance, without any rotation, reflection). However, human bodies are non-rigid objects and therefore the real situation is different obviously. Therefore, we argue that more accurate predictions of joints will lead to less residual, thus achieving a higher compression rate.

For inter prediction scheme, the corresponding skeletons in $(t-1)^{th}, (t-2)^{th}$ frames are used to predict the skeleton in $t^{th}$ frame (light yellow joints and dash lines) as shown in Fig.~\ref{fig:pre_fwk}. Then the differences between the original skeleton and the predicted skeleton are encoded.

\begin{figure*}[t]
\centering
\includegraphics[width=0.9\linewidth]{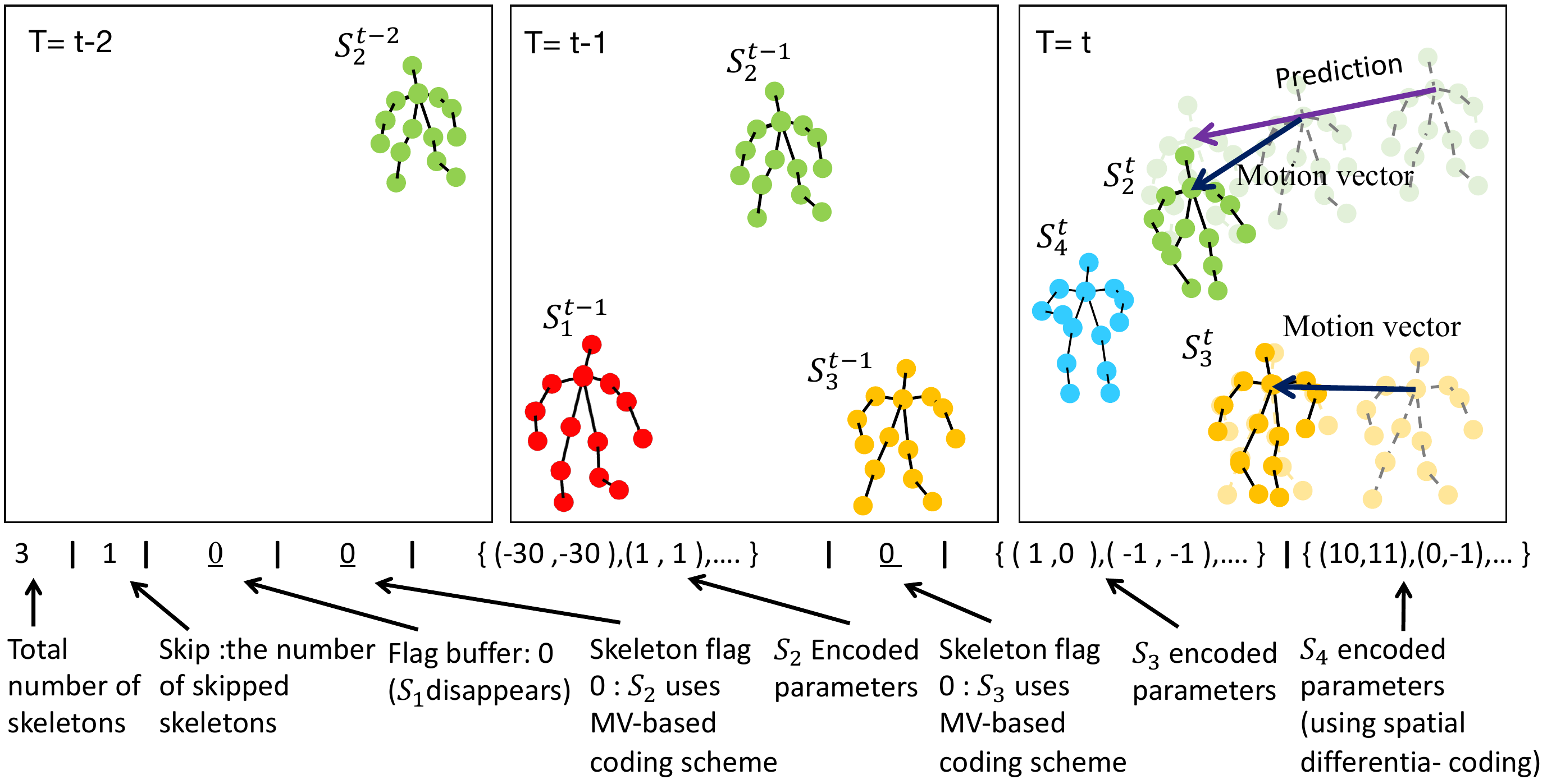}
\caption{An example of encoding skeletons in a frame with our multimodal coding tool.}
\label{fig:pre_example}
\end{figure*}

\textit{Trajectories prediction.}
There are a lot of researches working on trajectories prediction~\cite{antonini2006discrete, yamaguchi2011you, fernando2018soft+, gupta2018social}. In our method, the trajectories prediction method proposed in~\cite{gupta2018social} is used. More specifically, every key joint of a skeleton in $t^{th}$ frame is predicted individually with the corresponding joint in $(t-1)^{th}$ and $(t-2)^{th}$ frame (i.e. the $(t-1)^{th}$ and $(t-2)^{th}$ frames are chosen as the reference frames).


\subsection{Multimodal skeleton coding}
Considering labeling the skeleton data is expensive, the skeletons in videos may be the data estimated by the existing skeleton estimation methods. However, these methods may introduce some unexpected skeleton trajectories (for example, lack of key joints, inaccurate matching, and tracking), which leads to the correlations between skeletons become more complex and a more robust and efficient algorithm is needed. To this end, we propose a multimodal skeleton coding method where three schemes are switched for encoding skeletons.

The framework of our multimodal skeleton coding scheme has been shown in Fig.~\ref{fig:multi_framework}. Moreover, the switching rules are defined as follow:
 \begin{enumerate}[nolistsep,noitemsep]
 \item For a skeleton that newly appears in the current frame, the spatial difference-coding scheme is used. Besides, the spatial differential-coding scheme is also used for the first frame.
 \item When both MV-based differential-coding scheme and inter prediction can be used simultaneously for a skeleton, the one with less encoded bit length is chosen. A flag indicating the chosen scheme is allocated and transmitted.
 \item For other skeletons that exist in $(t-1)^{th}$ and $t^{th}$ but can not be found in $(t-2)^{th}$ frame, MV-based differential-coding scheme is used.
\end{enumerate}

\begin{figure}[t]
\centering
\includegraphics[width=\linewidth]{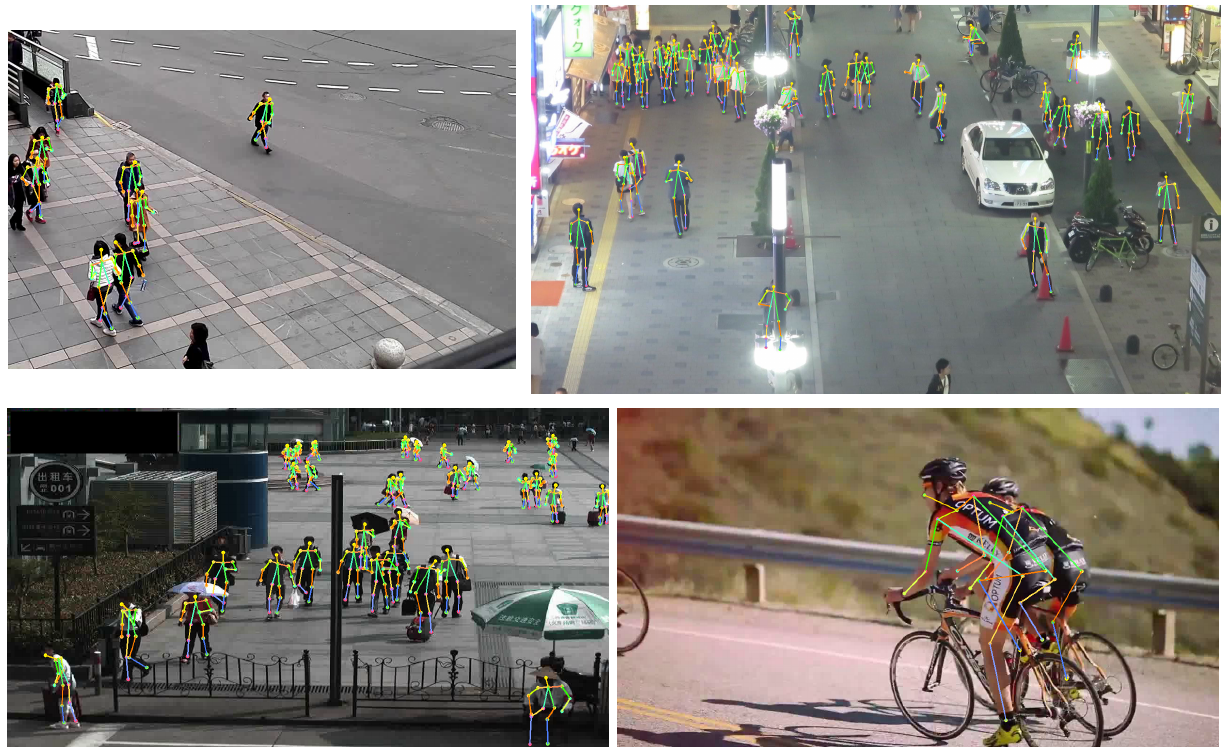}
\caption{Some example video frames}
\label{fig:snapshot}
\end{figure}

\begin{table*}[ht]
\centering
\caption{Experimental results of different coding schemes. Sequences 0, 1, 2 come from to PoseTrack dataset~\cite{PoseTrack}.}
\label{table1}
\setlength\tabcolsep{1.6pt}
\begin{threeparttable}
\begin{tabular}{|c|c|c|c|C{9mm}|C{14mm}|m{10mm}|c|c|c|c|}
\hline
 & & & & & & \multicolumn{5}{c|}{Size(KB)} \\
\cline{7-11}
\multirow{-2}{*}{Seq.}& \multirow{-2}{*}{Frames} & \multirow{-2}{*}{Resolution} & \multirow{-2}{*}{\#/Frame}& \multirow{-2}{\hsize}{Frame Skip} & \multirow{-2}{\hsize}{Skeletons Source}& Direct-coding & CM1& CM2 & CM3 & CM4 \\
\hline
 & &  &  &  & GT &  3.61 & 3.36(-6.7\%) & 0.86(-76.2\%) & 0.80(-77.8\%) & \textbf{0.78(-78.4\%)}  \\
& & &  & \multirow{-2}{*}{0} & ES & 3.65 & 3.26(-10.6\%) & \textbf{1.42(-61.2\%)} & 1.71(-53.1\%) & 1.42(-60.9\%)    \\
\cline{5-11}
& & &  & & GT &  1.86 & 1.74(-6.7\%) & 0.58(-68.8\%) & 0.55(-70.5\%) & \textbf{0.53(-71.5\%)}    \\

\multirow{-4}{*}{0}& \multirow{-4}{*}{31} & \multirow{-4}{*}{1280x720}& \multirow{-4}{*}{3} &\multirow{-2}{*}{1} & ES &  1.90 & 1.69(-11.3\%) & 0.90(-52.8\%) & 0.99(-47.9\%) & \textbf{0.86(-54.6\%)}    \\
\hline
& &  &  &  & GT &  2.42 & 2.26(-6.8\%) & 1.14(-52.9\%) & 1.23(-49.2\%) & \textbf{1.11(-53.9\%)}  \\
& &  &  & \multirow{-2}{*}{0} & ES & 2.42 & 2.22(-8.1\%) & \textbf{1.00(-58.6\%)} & 1.25(-48.2\%) & 1.01(-58.4\%)    \\
\cline{5-11}
& & &  & & GT &  1.25 & 1.16(-7.3\%) & \textbf{0.67(-46.5\%)} & 0.82(-34.3\%) & 0.67(-46.3\%)     \\
\multirow{-4}{*}{1}& \multirow{-4}{*}{31}& \multirow{-4}{*}{1280x720} & \multirow{-4}{*}{2}& \multirow{-2}{*}{1} & ES & 1.25 & 1.15(-8.3\%) & \textbf{0.61(-50.9\%)} & 0.86(-31.0\%) & 0.61(-50.8\%) \\
\hline
\multirow{4}{*}{2} & \multirow{4}{*}{31} & \multirow{4}{*}{1280x720} & \multirow{4}{*}{2} & & GT & 2.42 & 2.87(18.6\%) & 1.44(-40.4\%) & 1.28(-47.2\%) & \textbf{1.24(-48.6\%)} \\
& & & & \multirow{-2}{*}{0}  & ES & 2.74 & 3.21(17.3\%) & 2.15(-21.2\%) & 2.35(-14.0\%) & \textbf{2.19(-20.1\%)} \\
\cline{5-11}
& & & &  & GT & 1.25 & 1.48(18.6\%) & 0.89(-28.8\%) & 1.08(-13.9\%) & \textbf{0.88(-29.2\%)} \\
& & & & \multirow{-2}{*}{1}  & ES & 1.33 & 1.58(18.6\%) & \textbf{1.13(-15.1\%)} & 1.36(-2.2\%) & 1.15(-13.4\%) \\
\hline
\multirow{4}{*}{3} & \multirow{4}{*}{50} & \multirow{4}{*}{1008x672} & \multirow{4}{*}{8-10} & & GT  & 18.15 & 14.30(-21.2\%) & 5.61(-69.1\%) & 9.51(-47.6\%) & \textbf{5.60(-69.1\%)}  \\
& & & & \multirow{-2}{*}{0}  & ES & 20.93 & 16.56(-20.9\%) & 11.64(-44.4\%) & 14.01(-33.0\%) & \textbf{11.37(-45.7\%)} \\
\cline{5-11}
& & & &  & GT & 9.83 & 7.16(-27.1\%) &\textbf{ 4.38(-55.5\%)} & 5.63(-42.7\%) & 4.45(-54.8\%) \\
& & & & \multirow{-2}{*}{1} & ES & 10.43 & 8.24(-21.0\%) & 6.56(-37.1\%) & 7.77(-25.5\%) & \textbf{6.44(-38.2\%)} \\
\hline
\multirow{4}{*}{4} & \multirow{4}{*}{86} & \multirow{4}{*}{800x608} & \multirow{4}{*}{18-22} & & GT & 65.28 & 46.03(-29.5\%) & 15.09(-76.9\%) & 24.97(-61.8\%) & \textbf{14.16(-78.3\%)}  \\
& & & & \multirow{-2}{*}{0}  & ES & 76.29 & 52.69(-30.9\%) & 36.90(-51.6\%) & 46.00(-39.7\%) & \textbf{33.48(-56.1\%)} \\
\cline{5-11}
& & & & & GT & 32.66 & 23.02(-29.5\%) & 10.68(-67.3\%) & 15.76(-51.7\%) & \textbf{10.17(-68.8\%)} \\
& & & & \multirow{-2}{*}{1} & ES & 38.20 & 26.38(-30.9\%) & 20.42(-46.6\%) & 24.81(-35.1\%) & \textbf{18.81(-50.8\%)} \\
\hline
\multirow{4}{*}{5} & \multirow{4}{*}{80} & \multirow{4}{*}{1280x720} & \multirow{4}{*}{23-33} & & GT & 86.27 & 61.69(-28.5\%) & \textbf{14.36(-83.4\%)} & 29.96(-65.3\%) & 21.27(-75.3\%) \\
& & & & \multirow{-2}{*}{0} & ES & 86.28 & 58.86(-31.8\%) & 61.65(-28.5\%) & 89.65(-3.9\%) & \textbf{56.46(-34.6\%)}\\
\cline{5-11}
& & & & & GT & 43.31 & 30.94(-28.6\%) & \textbf{10.90(-74.8\%)} & 19.21(-55.6\%) & 13.98(-67.7\%) \\
& & & & \multirow{-2}{*}{1} & ES & 41.47 & 27.87(-32.8\%) & 29.56(-28.7\%) & 43.89(-5.8\%) & \textbf{27.76(-33.1\%)} \\
\hline
\multirow{4}{*}{6} & \multirow{4}{*}{100} & \multirow{4}{*}{1920x1080} & \multirow{4}{*}{34-35} &  & GT & 149.02 & 118.11(-20.7\%) & \textbf{7.43(-95.0\%)} & 16.95(-88.6\%) & 11.66(-92.2\%)\\
& & & & \multirow{-2}{*}{0}  & ES & 146.37 & 116.71(-20.3\%) & 85.42(-41.6\%) & 112.72(-23.0\%) & \textbf{77.44(-47.1\%)} \\
\cline{5-11}
& & & & & GT  & 86.24 & 52.08(-39.6\%) & \textbf{6.09(-92.9\%)} & 14.06(-83.7\%) & 9.67(-88.8\%) \\
& & & & \multirow{-2}{*}{1} & ES  & 71.25 & 56.52(-20.7\%) & 43.80(-38.5\%) & 58.05(-18.5\%) & \textbf{39.75(-44.2\%)}\\
\hline
\multicolumn{5}{|c|}{\multirow{2}{*}{Average on our surveillance seq.}} & GT& - & -28.1\% & \textbf{-76.9\%} & -62.1\% & -74.4\%  \\
\multicolumn{5}{|c|}{} & ES& - & -26.2\% & -39.6\% & -20.6\%  & \textbf{-43.7\%} \\

\hline

\multicolumn{6}{|c|}{\textbf{Average}} & - & -15.3\% & -53.8\% & -41.0\%  & \textbf{-54.7\%}\\
\hline
\end{tabular}
\end{threeparttable}
\end{table*}
Furthermore, several details should be noted: (1) For a skeleton that exists in the previous frame but disappears in the current frame, a disappear flag is allocated in bitstream. (2) For a skeleton that is exactly the same as its corresponding skeleton in the reference frame, a skip flag is allocated to indicate such condition instead of encoding fourteen zeros.

Fig~\ref{fig:pre_example} shows an example of coding skeletons in a frame using our proposed multimodal coding method. $S_{1}$ exists in all three frames and therefore both MV-based scheme and inter prediction scheme can be used. Finally, the MV-based scheme that leads to less bit length for encoding this skeleton is chosen and a flag is transmitted. Because $S_{2}$ only exists in the last and current frame, MV-based scheme is chosen. $S_{1}$ disappears in the current frame so that a skip flag is allocated. As for $S_{4}$ that newly appears in the current frame, the spatial differential-coding scheme is applied. The resulting bitstream of $t^{th}$ frame is also shown in Fig~\ref{fig:pre_example}.

\section{Experimental results}
\subsection{Settings}
In our experiments, the aforementioned four schemes (three single-modal schemes and one multimodal scheme) are evaluated and compared.

During the test, 7 videos with different resolutions and scenes are included. Three of them come from PoseTrack dataset~\cite{PoseTrack} and others are collected and labeled by ourselves. Some examples of them are shown in Fig.~\ref{fig:snapshot}. To evaluate the performance of our methods under different motion degrees, test sequences are re-sampled with different sample rates before being encoded. Apart from encoding the ground truth of skeletons (GT), we also evaluate our methods with skeletons estimated by~\cite{xiu2018pose} (ES). Note that only compression rate is used to evaluate our proposed lossless compression method.

\subsection{Results of different coding schemes.}
\label{sec:ex}
Table \ref{table1} compares the performance of different coding methods. In Table \ref{table1}, \emph{CM1} represents using the spatial differential-coding scheme; \emph{CM2} represents using the MV-based differential-coding; \emph{CM3} represents using the inter prediction scheme; \emph{CM4} represents our full version, multimodal coding method. Note that for a skeleton that MV-based scheme (inter prediction scheme) can not be used, spatial differential-coding scheme is used in \emph{CM2} (\emph{CM3}). From Table \ref{table1}, we can have the following observations:

\begin{enumerate}[nolistsep,noitemsep]
\item The full version of our approach, the multimodal coding method (\emph{CM4}), achieves the best performance on average. Specifically, it can reduce 54.7\% size of encoded skeleton data on average.

\item More importantly, our multimodal scheme shows superior performance (extra 4.1\% compression) to MV-based scheme when compressing estimated skeletons of surveillance sequences (i.e. the most practical situation). This demonstrates that our multimodal coding method is more robust than other compared methods when the skeletons trajectories in videos are complex and noisy and therefore is especially useful in the real applications.

\item When looking at encoding annotated skeletons of our collected surveillance sequences, 76.9\% and 74.4\% reduction of encoded size are obtained by our MV-based differential-coding scheme and multimodal coding method, respectively. This clearly indicates the effectiveness of our designed skeleton coding schemes.

\item Our MV-based scheme achieves 53.7\% compression rate across all test sequences, which is slightly worse than our multimodal scheme. This indicates that MV-based scheme can also provide satisfactory results at different kinds of applications.
\end{enumerate}

\section{Conclusion}
\label{sec:conc}
This paper presents a new skeleton coding tool for encoding skeletons in videos. We introduce a multimodal scheme where three encoding sub-schemes that utilize both spatial and temporal redundancy to compress skeleton data are switched properly, hence achieving higher coding efficiency. Experimental results show that skeleton data can be reduced efficiently using our multimodal coding tool.
\section *{\large Acknowledgement}

This paper is supported in part by: Shanghai ``The Belt and Road'' Young Scholar Exchange Grant (17510740100), the PKU-NTU Joint Research Institute (JRI) sponsored by a donation from the Ng Teng Fong Charitable Foundation.

\bibliographystyle{IEEEbib}
\bibliography{icme2019template}

\end{document}